\pdfoutput=1

\documentclass{ws-mpla}
\usepackage{graphicx}
\usepackage[super,compress]{cite}
\usepackage[breaklinks]{hyperref}
\hypersetup{colorlinks,urlcolor=black,citecolor=black,linkcolor=black,filecolor=black}
\usepackage{breakurl}
\usepackage[dvipsnames]{xcolor}

\graphicspath{{./fig/}}
\newcommand{\dcc}{LIGO-P2100017}

\newcommand{\pyRingFrequencyDeviationPop}{$ \delta \hat{f}_{221} = 0.04^{+ 0.27 }_{- 0.32 } $ }

\newcommand{\pSEOBFrequencyDeviationPop}{$\delta \hat{f}_{220} = 0.03^{+ 0.38 }_{- 0.35 } $ }

\begin{document}

\markboth{M.~Isi}
{Recent LIGO-Virgo discoveries}

\catchline{}{}{}{}{}

\title{Recent LIGO-Virgo discoveries}

\author{\footnotesize Maximiliano Isi\footnote{
NHFP Einstein fellow}}

\address{LIGO Laboratory, Massachusetts Institute of Technology\\
Cambridge, Massachusetts 02139, USA\\
maxisi@mit.edu}

\maketitle

\pub{Received (Day Month Year)}{Revised (Day Month Year)}

\begin{abstract}
The LIGO and Virgo gravitational-wave detectors carried out the first half of their third observing run from April through October 2019.
During this period, they detected 39 new signals from the coalescence of black holes or neutron stars, more than quadrupling the total number of detected events.
These detections included some unprecedented sources, like a pair of black holes with unequal masses (GW190412), a massive pair of neutron stars (GW190425), a black hole potentially in the supernova pair-instability mass gap (GW190521), and either the lightest black hole or the heaviest neutron star known to date (GW190814).
Collectively, the full set of signals provided astrophysically valuable information about the distributions of compact objects and their evolution throughout cosmic history.
It also enabled more constraining and diverse tests of general relativity, including new probes of the fundamental nature of black holes.
This review summarizes the highlights of these results and their implications.

\end{abstract}

\ccode{PACS Nos.: include PACS Nos.}

\section{Introduction}

Gravitational waves (GWs) are self-propagating perturbations in the curvature of space-time, originating in the acceleration of massive bodies.
They manifest locally by alternatively stretching and squeezing distances along axes perpendicular to their direction of propagation.
Such minute strains can be recorded by exquisitely precise detectors on Earth, like LIGO\cite{TheLIGOScientific:2014jea} and Virgo\cite{TheVirgo:2014hva}, and used to extract valuable information about the violent astrophysical processes that produced them.

Although predicted by Einstein in 1916, GWs were not directly detected until the two Advanced LIGO instruments came online in September, 2015.\cite{GW150914_paper,TheLIGOScientific:2016agk}
That historic first signal, GW150914, originated from the collision of a pair of black holes $1.3~\mathrm{Gly}$ away from Earth.\cite{TheLIGOScientific:2016wfe}
LIGO further observed two other such binary black hole (BBH) coalescences during its first observing run (O1),\cite{GW151226,TheLIGOScientific:2016pea} which ended in January, 2016; it detected seven more during its second observing run (O2), which took place from the end of November, 2016 through August, 2017, by which point LIGO had been joined by Virgo.\cite{GWTC1}
Besides BBHs, O2 also brought the first binary neutron star (BNS) coalescence,\cite{TheLIGOScientific:2017qsa} which was detected not only through GWs but also electromagnetic radiation across the spectrum, delivering a wealth of astrophysical information.\cite{Monitor:2017mdv,GBM:2017lvd}

This review focuses on the LIGO-Virgo discoveries since 2017.
Section \ref{sec:det} summarizes the detections made in the first part of the third observing run (O3a) as presented in the second LIGO-Virgo catalog (GWTC-2),\cite{GWTC2} with subsections dedicated to the exceptional events GW190412, GW190425, GW190521, and GW190814.\cite{GW190412,GW190425,GW190521g,GW190521g:imp,GW190814}
Section \ref{sec:astro} outlines astrophysical implications.
Section \ref{sec:tgr} summarizes tests of general relativity.
Section \ref{sec:conclusion} provides concluding remarks.

\section{Detections}
\label{sec:det}

\begin{figure}
\centerline{\includegraphics[width=0.7\textwidth]{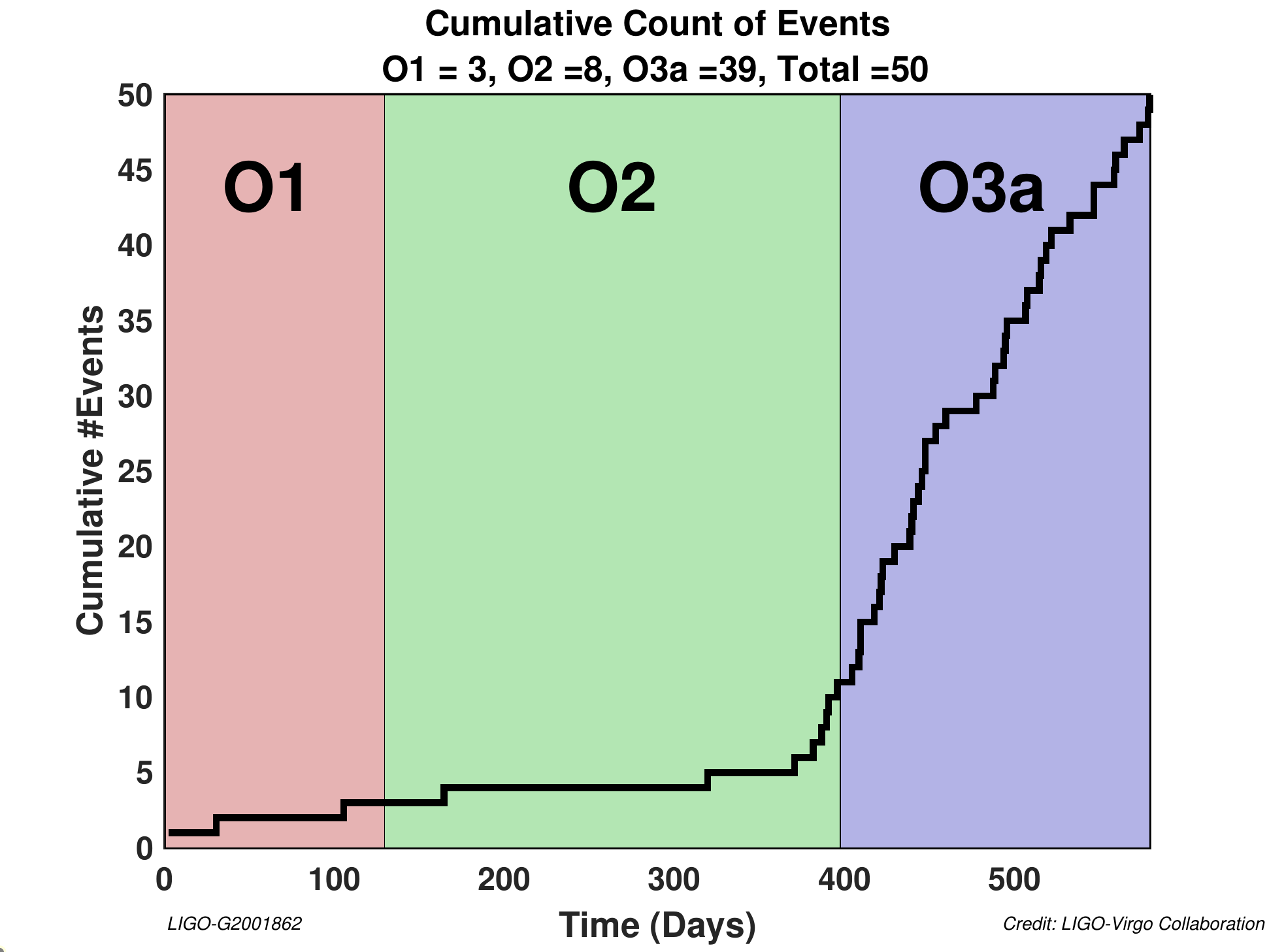}}
\vspace*{8pt}
\caption{Cumulative number of LIGO-Virgo detections up to the end of O3a.\cite{G2001862}}
\label{fig:detections}
\end{figure}

After a downtime period for commissioning and instrumental upgrades, LIGO and Virgo began O3a on April 1, 2019, and continued observing until October 1, 2019.\cite{GWTC2}
Enhancements to the instruments led to an average ${\sim}60\%$ improvement in the detection range with respect to O2, as well as an increase in the network duty cycle.\cite{Buikema:2020dlj}
Both these factors resulted in a significantly greater rate of detections, as shown in Fig.~\ref{fig:detections}.
Throughout O3a, LIGO and Virgo detected 39 compact-binary coalescences (CBCs) with a false-alarm rate (FAR) smaller than 2 per year.

Out of the 39 new triggers in GWTC-2, 36 can be attributed to compact binaries in which both components are heavier than $3\, M_\odot$ and, therefore, can be confidently categorized as BBHs.
The remaining three triggers show signs that at least one of its source components may have been a neutron star (NS), which must be lighter than ${\sim}3\, M_\odot$.
The potentially non-BBH candidates are: GW190425, which is most likely a BNS; GW190814, which originated from either a BBH or a neutron-star--black-hole (NSBH) merger; and GW190426\_152155, the candidate with the highest FAR in GWTC-2, which could also correspond to either a BBH or an NSBH.\cite{GWTC2}

Out of the multiple BBHs detected in O3a, GW190412 and GW190521 received special attention because of their unusual mass ratio and total mass, respectively; similarly, as the most significant potentially-non-BBH triggers, GW190425 and GW190814 also received particular focus.
All of these four detections were treated in dedicated publications,\cite{GW190412,GW190425,GW190521g,GW190521g:imp,GW190814} and are summarized chronologically in the subsections below.
The implications of the full population of detections are fleshed out in the following two sections.

\subsection{GW190412}
\label{sec:gw190412}

All systems detected during O1 and O2 were consistent with binaries in which both components had equal masses.\cite{GWTC1,LIGOScientific:2018jsj,Roulet:2018jbe}
This changed with the detection of GW190412,\cite{GW190412} which pointed to black holes with masses $m_1 = 30.1^{+4.6}_{-5.3}\, M_\odot$ and $m_2 = 8.3^{+1.6}_{-0.9}\, M_\odot$, for a mass ratio $q = m_2/m_1 =  0.28^{+0.12}_{-0.07}$ inconsistent with unity.%
\footnote{Unless otherwise noted, all quantities reported in this review consist of the median and the symmetric $90\%$-credible interval around it.}
The source of GW190412 was also the third binary in which LIGO-Virgo identified at least one spinning component, with an effective spin parameter $\chi_\mathrm{eff} = 0.25^{+0.08}_{-0.11}$.
Unlike for the previous detections, GW190412's unequal masses allowed us to unambiguously assign some of this angular momentum to the more massive component, whose dimensionless spin was pinned down to $\chi_1 = 0.44^{+0.16}_{-0.22}$; on the other hand, the spin of the lighter component was largely unconstrained.
The atypical mass ratio and spin of the GW190412 binary have motivated several studies on its possible origins.\cite{Rodriguez:2020viw,Safarzadeh:2020qrc,Gerosa:2020bjb,Hamers:2020huo,Olejak:2020oel,Mandel:2020lhv}

GW190412 was also interesting due to its signal morphology.
The mass ratio and inclination of the source resulted in detectable contributions from the higher multipoles of the signal, that is, GW radiation modes with angular number $\ell \geq 3$.\cite{GW190412}
This was a first, since only the $\ell = |m| = 2$ multipole had been identified in previous detections (and the majority of detections since).
Unfortunately, this signal was also the first to present significant differences across waveform models, highlighting the importance of further waveform development in this corner of parameter space.

\subsection{GW190425}
\label{sec:gw190425}

GW190425 was the second likely BNS detected by LIGO-Virgo,\cite{GW190425} following GW170817.\cite{TheLIGOScientific:2017qsa}
Unlike for that first BNS detection, however, there was no confirmed electromagnetic counterpart associated with this GW trigger (although see Ref.~[\protect\refcite{Pozanenko:2019lwh}]).
Although both LIGO Livingston and Virgo were collecting data at the time, GW190425's network signal-to-noise ratio (SNR) of 12.9 was overwhelmingly dominated by the Livingston detector alone, resulting in a poor sky localization ($8285~\mathrm{deg}^2$ at 90\% credibility).
The GW data imply source component masses in the range $1.12$ to $3.52\, M_\odot$, which is consistent with a pair of neutron stars; however, there is no evidence of tidal deformability, or any other matter effects that would be necessary to unequivocally rule out the BBH hypothesis.

Assuming GW190425 did originate in a BNS, its source mass of $3.4^{+0.3}_{-0.1}\, M_\odot$ would make it stand out in relation to the known binary pulsars in the Milky Way, the heaviest of which has a total mass of $2.89\, M_\odot$.\cite{Farrow:2019xnc}
This fact provides a clue about the origin of the GW190425 binary, which could have formed from a tight stellar binary progenitor, or through a dynamical encounter.\cite{GW190425,Safarzadeh:2020efa,Romero-Shaw:2020aaj}
The alternative hypotheses that one or both of the GW190425 binary components are actually black holes is inconsistent with traditional formation channels.\cite{GW190425}

\subsection{GW190521}
\label{sec:gw190521}

\begin{figure}
\centerline{\includegraphics[width=\textwidth]{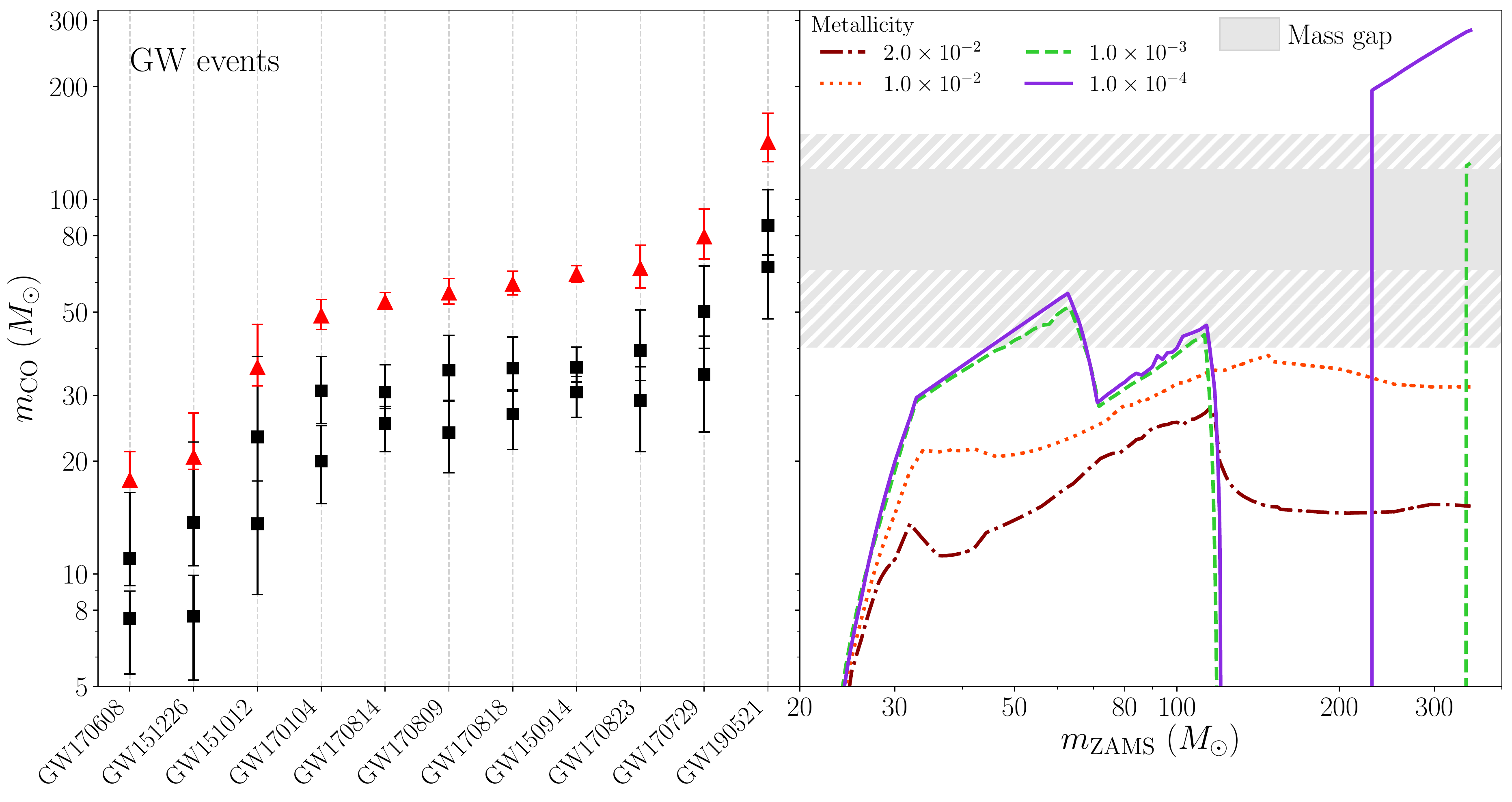}}
\vspace*{8pt}
\caption{\textit{Left:} Component (black) and remnant (red) masses of BBHs up to GW190521. \textit{Right:} Mass of remnant compact object ($m_{\rm CO}$) as a function of its the zero-age main-sequence mass of the progenitor star ($m_{\rm ZAMS}$) according to theoretical models for star evolution with different metalicities (color);\cite{Bressan:2012zf,Fryer:2011cx,Woosley:2016hmi} the shaded region higlights the mass gap, and hatching marks the uncertainty in its boundaries.
(Figure reproduced from Ref.~[\protect\refcite{GW190521g:imp}]; see that work for details.)
}
\label{fig:gw190521}
\end{figure}

GW190521 is arguably the most intriguing of all BBH detections in O3a.
The main reason for this is that, with $m_1 = 85^{+21}_{-14}\, M_\odot$,%
\footnote{Values quoted here were obtained with the \textsc{NRSur7dq4} waveform model, as in Ref.~[\refcite{GW190521g}].}
the binary's heaviest component falls squarely within the supernova pair-instability mass gap, a range of black hole masses that cannot be easily explained.\cite{GW190521g}
The gap is expected to arise due to nuclear processes that disrupt the cores of massive stars towards the end of their lives, preventing stellar progenitors from forming black holes with masses in the approximate range ${\sim}65{-}120\,M_\odot$.\cite{Woosley:2016hmi}
Even though the predicted boundaries for this range vary, the gap itself is a robust prediction of numerical simulations, with $65\,M_\odot$ a conservative lower limit.\cite{Woosley:2016hmi,2019ApJ...878...49W,Heger:2002by,Belczynski:2016jno,Spera:2017fyx,Giacobbo:2017qhh,Farmer:2019jed,Marchant:2018kun,Stevenson:2019rcw,Mapelli:2019ipt}
There is only a 0.32\% probability for the mass of the GW190521 primary to have been below the $65\, M_\odot$ boundary.\cite{GW190521g:imp}
The formation of this object thus calls for alternative explanations, like a second-generation merger, i.e., a black hole formed from the merger of two smaller black-hole progenitors.
If such a merger occurred in the accretion disk of an active galactic nucleus, it could have resulted in an electromagnetic counterpart---a tantalizing possibility explored in Ref.~[\protect\refcite{Graham:2020gwr}].

The secondary component of the GW190521 binary was inferred to have $m_2 = 66^{+0.28}_{-0.34}\, M_\odot$, which means that this was the heaviest system detected by LIGO-Virgo to date, with a total mass $M = m_1 + m_2 = 150^{+29}_{-17}\, M_\odot$.
The remnant formed in this merger had a correspondingly large mass of $m_f = 142^{+28}_{-16}\, M_\odot$, making it the first confirmed intermediate-mass black hole (IMBH), i.e., a hole in the range $10^2{-}10^5\, M_\odot$.
This category is of interest as a potential bridge between stellar-mass black holes ($\lesssim 10^2\, M_\odot$) and the supermassive black holes at the centers of galaxies ($\gtrsim 10^5\, M_\odot$).\cite{Volonteri:2010wz,Greene:2019vlv}

Its high source mass makes GW190521 a challenging signal to analyze: there is hardly (if any) detectable power in the inspiral, which occurs at frequencies too low for the LIGO-Virgo instruments to properly detect.
Because the signal is dominated by the merger-ringdown, it is not possible to unequivocally establish that its source was a quasicircular black-hole coalescence, as is standard to assume, and not some other less astrophysically plausible scenario, like an eccentric merger.\cite{Romero-Shaw:2020thy,Gayathri:2020coq}
Under the assumption that this was indeed a quasicircular BBH, the data seem to indicate that the component spins were high, and that the plane of the orbit may have precessed due to a misalignment of the spins with respect to the orbital angular momentum;\cite{GW190521g:imp} although the evidence for such effect is weak, it has added to the interest in this event.\cite{Nitz:2020mga,Shibata:2021sau,Cruz-Osorio:2021qbr,Tanikawa:2020abs,Rice:2020gyx,Liu:2020lmi,Safarzadeh:2020vbv,Liu:2020gif,Farrell:2020zju,Fishbach:2020qag,Graham:2020gwr}

\subsection{GW190814}
\label{sec:gw190814}

Besides the supernova pair instability mass gap, it has been suggested that a second, lower mass gap could exist in the range $2.5{-}5\, M_\odot$, separating neutron stars from black holes.\cite{Bailyn:1997xt,Farr:2010tu,Ozel:2010su,Ozel:2012ax}
The detection of GW190814 seems to invalidate this hypothesis: with a mass $m_2 = 2.59^{+0.08}_{-0.09}\, M_\odot$, the secondary object in the GW190814 binary falls right in this lower gap, meaning that this object was either the heaviest neutron star or the lightest black hole ever observed.\cite{GW190814}
On the other hand, the primary object in GW190814 had a mass $m_1 = 23.2^{+1.1}_{-1.0}\, M_\odot$, which is fully consistent with a black hole, and makes GW190814 a potential NSBH detection.

Yet, the NSBH hypothesis presents important challenges.
Without signs of tidal effects in the GW data, and in the absence of an electromagnetic counterpart, there is no clear evidence that the lightest object is indeed a neutron star.\cite{GW190814}
Furthermore, it would be difficult to account for the relatively high mass of the putative neutron star in the GW190814 binary: although some models for the nuclear equation of state could allow for masses up to ${\sim}3\,M_\odot$, these are in tension with the measurements from GW170817, which bound the maximum neutron star mass to be $\lesssim 2.4\, M_\odot$,\cite{Abbott:2018wiz,Lim:2019som,Essick:2019ldf,Abbott:2018exr} as well as with the observed population of galactic pulsars, the heaviest of which has a mass of $2.14^{+0.10}_{-0.09}\,M_\odot$ within $68\%$ confidence.\cite{Cromartie:2019kug}
Modulo the caveats implied by their respective modeling assumptions, these considerations strongly disfavor the NSBH scenario for GW190814.

On the other hand, the alternative that GW190814 originated from a BBH raises its own, albeit less daunting, questions.
The reason for that lies primarily in the inferred mass ratio $q = 0.112^{+0.008}_{-0.008}$, which astrophysical simulations do not tend to produce in sufficiently high numbers to explain this detection.\cite{GW190814}
This is true for a variety of formation channels, including isolated binaries\cite{Dominik:2012kk,Dominik:2014yma,Marchant:2017ehn,Giacobbo:2018etu,Kruckow:2018slo,Neijssel:2019irh,Olejak:2020oel} and dynamical encounters in globular clusters.\cite{Sigurdsson:1993zrm,Rodriguez:2016kxx} Alternative environments, like young star clusters\cite{Ziosi:2014sra,DiCarlo:2019pmf,Rastello:2020sru,Sedda:2021oov} or the disks surrounding active galactic nuclei,\cite{Yang:2019okq,McKernan:2020lgr} might prove more suitable.
Either way, GW190814 provides an important clue for understanding the formation of compact binaries, and will doubtlessly inform population models.

\section{Astrophysical implications}
\label{sec:astro}

\begin{figure}
\centerline{\includegraphics[width=0.5\textwidth]{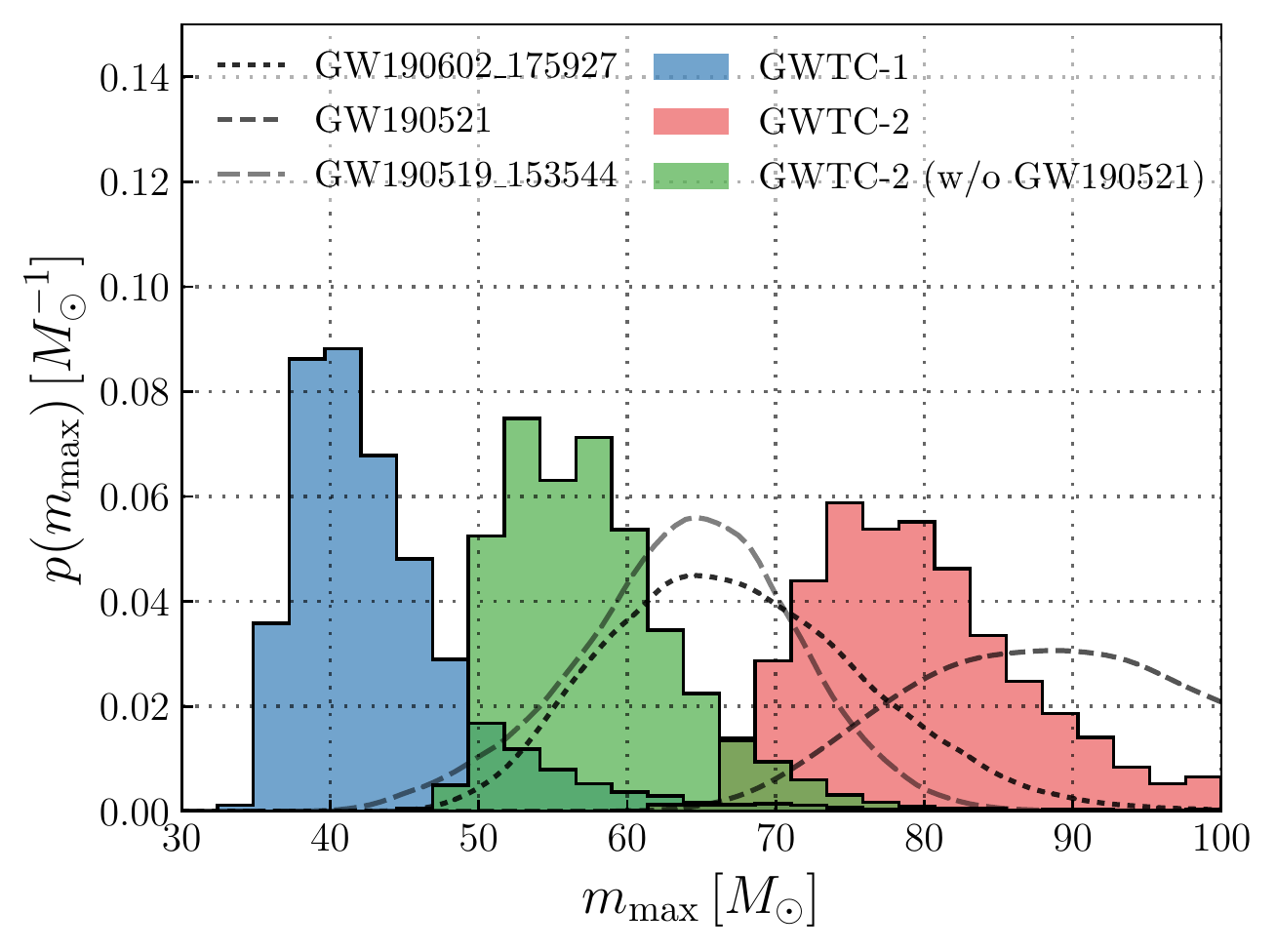}
\includegraphics[width=0.5\textwidth]{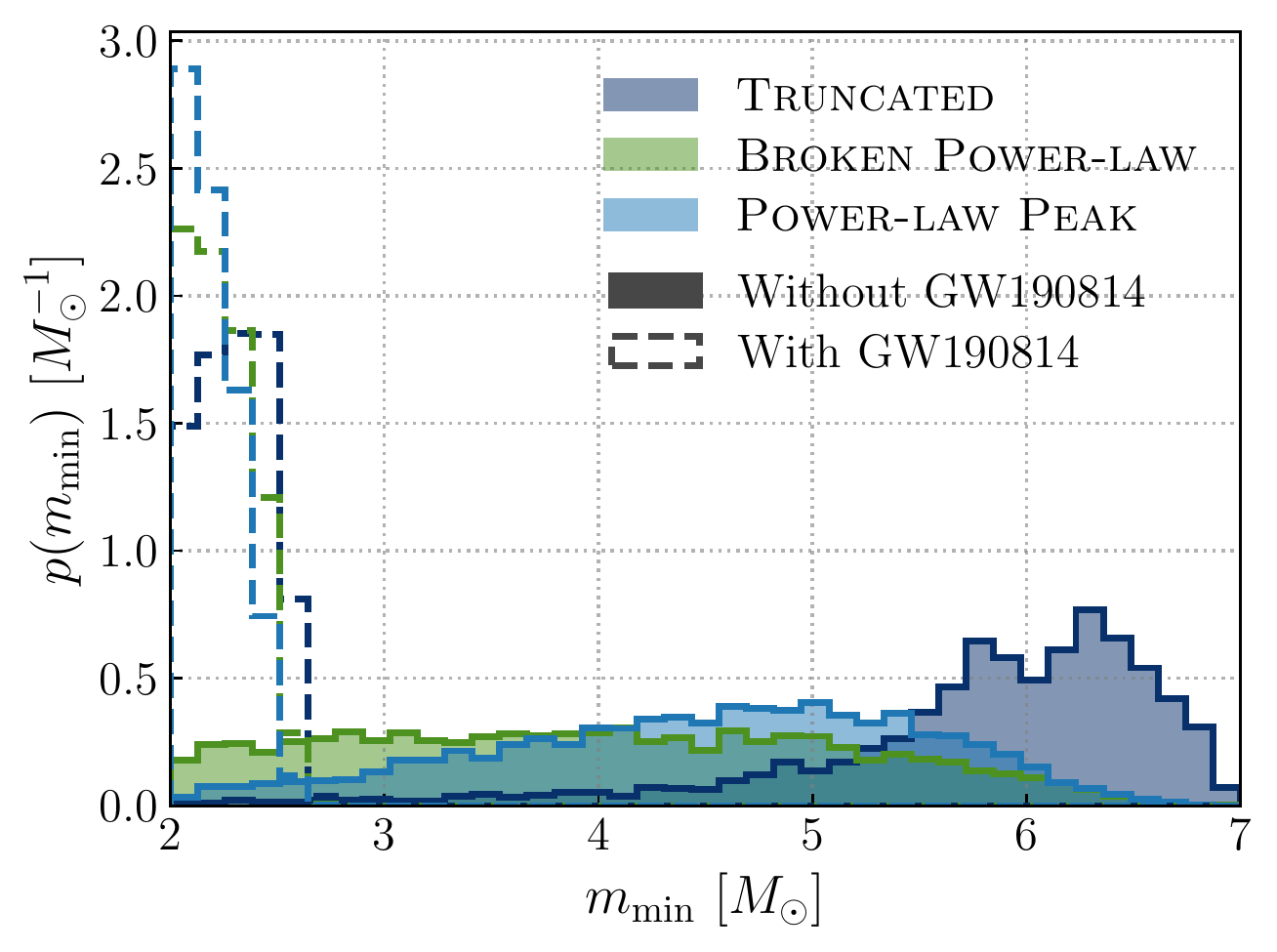}}
\vspace*{8pt}
\caption{
\textit{Left:} primary BBH mass ($m_{\rm max}$) posterior obtained from GWTC-1 (blue), GWTC-2 (red) and GWTC-2 without GW190521 (green), compared to the posteriors of some individual events; GW190521 is in tension with the GWTC-1 population, but not GWTC-2.
\textit{Right:} secondary BBH mass ($m_{\rm min}$) posterior obtained with different population models for the GWTC-2 BBH mass spectrum (color), with (dashed) and without (solid) including GW190814 in the mix; the GW190814 secondary mass is an outlier of the GWTC-2 BBH population.
(Figures reproduced from Ref.~[\protect\refcite{Abbott:2020gyp}]; see that work for details.)
}
\label{fig:astro}
\end{figure}

The collection of LIGO-Virgo detections can be studied as a whole to learn about the astrophysical population of merging black holes and neutron stars.\cite{LIGOScientific:2018jsj,Roulet:2018jbe,Abbott:2020gyp}
Modeling their mass and spin distributions, one can draw conclusions about merger rates, formation channels, and their evolution through cosmic history.
Typically, this is achieved through hierarchical (also known as ``extreme deconvolution''\cite{Bovy:2009kp}) Bayesian models that can be used to measure parameters describing a population of events, rather than isolated GW sources.\cite{Loredo:2004nn,Mandel:2009pc,Mandel:2009nx}
Where applicable, such multilevel inference models can also be made to account for selection biases, by which sources with some sets of parameters are more easily detected than others.\cite{Mandel:2018mve,Thrane:2018qnx,Vitale:2020aaz}
Doing so is necessary to extract astrophysical information from LIGO-Virgo data.

Back in 2018, the detections in the first LIGO-Virgo catalog (GWTC-1)\cite{GWTC1} had already provided some insights on the merging BBH and BNS populations.\cite{LIGOScientific:2018jsj}
Crucially, GWTC-1 suggested that only a minority ($<1\%$) of BBHs involve a component with mass greater than $45\, M_\odot$, and that it is uncommon for such systems to have large component spins aligned with the orbital angular momentum.
The BBH and BNS merger rates were respectively constrained between $25{-}109~\mathrm{Gpc}^{-3} \mathrm{yr}^{-1}$ and $110{–}2520~\mathrm{Gpc}^{-3} \mathrm{yr}^{-1}$; there was a strong indication that the BBH merger rate does not decrease with redshift.

The additional detections in GWTC-2 add further complexity to this picture.\cite{Abbott:2020gyp}
Unlike in GWTC-1, the BBH mass distribution in GWTC-2 is no longer well-described by a simple power law with a high-mass cutoff, instead hinting at the presence of some nontrivial structure (like a peak, or a change in power index) around ${\sim}30{-}40\,M_\odot$.
Notably, GW190521, the most massive system detected so far, is in tension with the GWTC-1 population, which disfavored BBHs with masses $> 45\, M_\odot$, but not with the GWTC-2 BBH population as a whole (Fig.~\ref{fig:astro}, left).
The same is true of GW190412, whose unusual mass ratio is consistent with the GWTC-2 models.
On the other hand, the low mass of the GW190814 secondary component is difficult to explain in the context of the GWTC-2 BBH detections (Fig.~\ref{fig:astro}, right).

Besides black hole masses, GWTC-2 also provides important insights about spins.
The BBH detections indicate with high credibility that some of the spins must be misaligned with the orbital angular momentum, leading to precession;
this is in agreement with previous conclusions drawn from GWTC-1 data.\cite{Farr:2017uvj,Tiwari:2017ndi,Biscoveanu:2020are}
Moreover, a nonnegligible portion of BBHs ($12{-}44\%$) are likely to contain spins tilted over $90^{\circ}$ with respect to the orbital angular momentum. 
These are all important clues because spin (mis)alignment may be reflective of a binary's formation mechanism: for example, common envelope evolution tends to to align the individual spins with the orbital angular momentum,\cite{Rodriguez:2016vmx,Stevenson:2017dlk,Gerosa:2018wbw,Bavera:2019fkg} while dynamical capture does not.\cite{Kalogera:1999tq,Mandel:2009nx,Rodriguez:2016vmx,Doctor:2019ruh}

Merger rate estimates improve with GWTC-2: the data constrain the rate to $23.9^{+14.3}_{-8.6}~\mathrm{Gpc}^{-3} \mathrm{yr}^{-1}$ for BBHs,\cite{Abbott:2020gyp} and $320^{+490}_{-240}~\mathrm{Gpc}^{-3} \mathrm{yr}^{-1}$ for BNSs.\cite{GW190425}
Furthermore, GWTC-2 confirms the expectation from GWTC-1 that the BBH rate is likely to increase with redshift, although the degree of this growth is now constrained with high credibility to be below that of the star formation rate.\cite{Abbott:2020gyp}

\section{Tests of general relativity}
\label{sec:tgr}

The inference of GW source properties is only possible thanks to the availability of waveform models that accurately represent the predictions of general relativity (GR) for a given system, like the quasicircular coalescence of two black holes or neutron stars.%
\footnote{The literature on GW waveforms is vast; examples for the most widely-used waveform models include Refs.~[\refcite{Bohe:2016gbl,Pan:2013rra,Babak:2016tgq,Ossokine:2020kjp,Cotesta:2018fcv,Cotesta:2020qhw,Husa:2015iqa,Khan:2015jqa,Hannam:2013oca,Khan:2018fmp,Khan:2019kot,Varma:2019csw}].}
Rather than taking these models for granted, we may ask about their overall agreement with the data, and look for evidence of modeling systematics\cite{Bohe:2016gbl,Khan:2015jqa,Cotesta:2018fcv,Varma:2018mmi,Cotesta:2020qhw,Ramos-Buades:2020noq,Pratten:2020fqn,Pratten:2020ceb,Ossokine:2020kjp} or, more crucially, deviations from GR itself.\cite{Will:2014kxa}
It is standard to refer loosely to studies of this kind as ``tests of GR,'' although it should always be understood that they are sensitive to systematic errors more broadly, not just fundamental physics: should these analyses turn up something interesting, further work would be required to discard the presence of ordinary systematic errors before concluding that a GR violation has been observed; if no statistically significant anomalies are found, one can assert there is evidence for neither modeling errors nor deviations from GR.

The first tests of GR with LIGO-Virgo signals were carried out with GW150914,\cite{TheLIGOScientific:2016src} laying the groundwork for many of the analyses that would be repeated for subsequent detections.\cite{TheLIGOScientific:2016pea,LIGOScientific:2019fpa,Abbott:2020jks}
This included waveform consistency tests, parameterized tests of the source dynamics, studies of GW dispersion, and probes of the merger remnant.
A study of GW polarizations was also attempted with GW150914, but the results were inconclusive; it was not until Virgo joined the network with GW170814 that some elementary polarization tests became viable.\cite{Abbott:2017oio}
Later on, GW170817 allowed for some unique analyses relying on the electromagnetic counterpart, including a measurement of the GW speed and tests for extra spatial dimensions, which have not been repeated since.\cite{Abbott:2018lct}
After that, GWTC-1 made it possible to combine results from multiple BBH events to obtain more stringent constraints on deviations from GR, although this was done under restrictive assumptions.\cite{LIGOScientific:2019fpa}

\begin{figure}
\centerline{\includegraphics[width=1.025\textwidth]{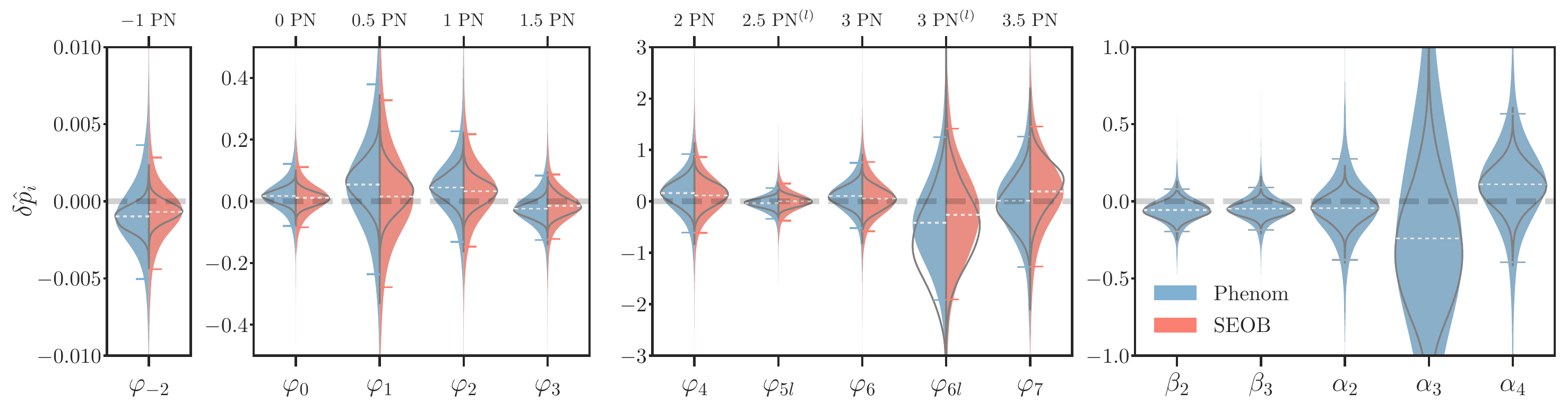}}
\vspace*{8pt}
\caption{
Measurement of PN (first through third panels) and merger-ringdown (fourth panel) coefficients from BBHs in GWTC-2.
Filled distributions represent the result from combining the individual measurements from each of the events through a hierarchical model, which does not require assuming that the coefficients take a single, shared value for all events; unfilled distributions represent the combined result assuming each coefficient takes a single value shared by all events, as done previously for GWTC-1.
Color corresponds to the baseline GR-based waveform.
(Figure reproduced from Ref.~[\protect\refcite{Abbott:2020jks}]; see that work for details.)
}
\label{fig:tgr}
\end{figure}

GWTC-2 enabled quantitative and qualitative improvements over previous results, leveraging the increased number of detections and the greater variety in their properties.\cite{Abbott:2020jks}
Due to computational constraints, LIGO-Virgo focused on the BBHs of GWTC-2 with a FAR better than $1/1000$ per year, a criterion satisfied by 24 signals.
These signals were analyzed individually and collectively through eight different tests, including a few novel additions to the LIGO-Virgo repertoire, like probes of the no-hair theorem and a search for echoes.
Ref.~[\refcite{Abbott:2020jks}] also took advantage of hierarchical modeling techniques to evaluate the population as a whole (as in Sec.~\ref{sec:astro}), doing away with some of the limitations affecting the GWTC-1 combined results.

Highlights from the GWTC-2 testing GR analyses include a factor of ${\sim}2$ improvement on constraints of post-Newtonian (PN) coefficients.\cite{Abbott:2020jks}
The PN coefficients control the details of the GW phase evolution and originate from a series expansion of the source dynamics in powers of $v/c$, where $v$ is the orbital velocity during the inspiral and $c$ is the speed of light: a coefficient of PN order $n=i/2$ corresponds to a source term with dependence $(v/c)^{i}$, and enters the GW phasing as a factor $f^{(i-5)/3}$ of the Fourier frequency $f$.\cite{Blanchet:2013haa}
In GR, these quantities are known analytically as a function of the source parameters (masses and spins, and potentially tidal deformability) up to high order,\cite{Blanchet:1995ez,Kidder:1995zr,Damour:2001bu,Blanchet:2005tk,Blanchet:2006gy,Arun:2008kb,Bohe:2012mr,Marsat:2012fn,Bohe:2013cla,Blanchet:2013haa,Damour:2014jta,Bohe:2015ana} but they can take different values in beyond-GR theories of gravity.\cite{Blanchet:1994ez,Blanchet:1994ex,Arun:2006hn,Arun:2006yw,Mishra:2010tp,Yunes:2009ke}
Although the PN parameters only refer to the inspiral regime, phenomenological coefficients can be defined to control deviations from GR during the merger and ringdown.\cite{Meidam:2017dgf}
Figure \ref{fig:tgr} shows GWTC-2 measurements of PN and phenomenological coefficients; all posteriors are consistent with GR and can be used to place constraints on specific theories beyond GR.\cite{Will:2014kxa,Yunes:2009ke,Yunes:2016jcc,Nair:2019iur}
Besides tightening GWTC-1 constraints and relaxing previous assumptions for combining events, GWTC-2 results also stand out due to the inclusion of higher order multipoles of the radiation for some of the events (GW190412, GW190521, and GW190814).

Thanks to the detection of heavier events, GWTC-2 also enabled studies targeting the merger remnant through the ringdown radiation,\cite{Carullo:2019flw,Isi:2019aib,Brito:2018rfr, Cotesta:2018fcv} as well as searches for potential postmerger echoes.\cite{Lo:2018sep}
Ringdown analyses model the end stage of each GW signal as a superposition of damped sinusoids, which in GR are expected to have specific frequencies and damping rates determined by the remnant mass and spin.\cite{Vishveshwara1970b, Press1971, Teukolsky:1973ha, Chandrasekhar:1975zza}
The frequencies and damping rates extracted from the data can be evaluated for consistency with a given mass and spin, as expected from the preceding inspiral or by comparing different ringdown modes.\cite{Detweiler:1980gk, Dreyer:2003bv, LISA_spectroscopy, Gossan:2011ha, PhysRevD.90.064009, Carullo:2018sfu, Brito:2018rfr, Carullo:2019flw, Isi:2019aib, Bhagwat:2019bwv, Bhagwat:2019dtm,Cabero:2019zyt}
Using GWTC-2, such spectroscopic measurements yielded constraints on the fractional deviation away from the GR frequencies of \pSEOBFrequencyDeviationPop for the fundamental $\ell=m=2$ ringdown mode, and \pyRingFrequencyDeviationPop for the its first overtone.\cite{Abbott:2020jks}
There was no evidence for echoes following the ringdown signal for any of the events analyzed.\cite{Abbott:2020jks}

\begin{figure}
\centerline{\includegraphics[width=0.7\textwidth]{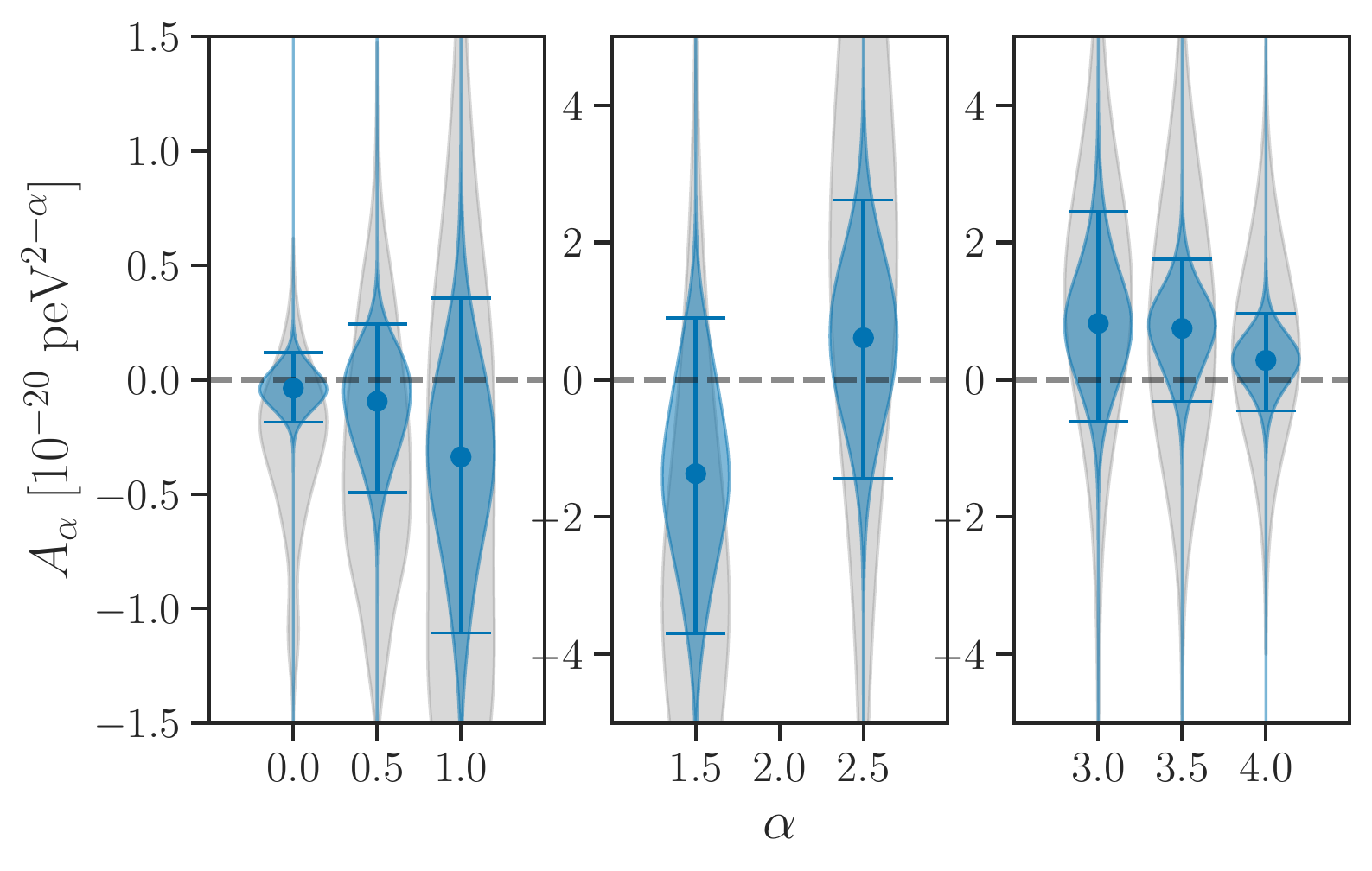}}
\vspace*{8pt}
\caption{
Measurement of coefficients controlling the GW dispersion from GWTC-2 (blue), compared to GWTC-1 (gray).
The dispersion relation is parameterized as $E^2 = p^2c^2 + A_\alpha p^\alpha c^\alpha$, with $E$ and $p$ respectively the graviton energy and momentum; the Lorentz-violating coefficients $A_\alpha$ all vanish in GR.
(Figure reproduced from Ref.~[\protect\refcite{Abbott:2020jks}]; see that work for details.)
}
\label{fig:liv}
\end{figure}

The GWTC-2 detections also produced the best limits yet on the presence of GW dispersion---an effect that is absent in GR but which can arise in multiple beyond-GR scenarios, including due to a massive graviton or generic Lorentz-violations.\cite{Mirshekari:2011yq,Yunes:2016jcc,Will:1997bb,Calcagni:2009kc,AmelinoCamelia:2002wr,
Horava:2009uw,Sefiedgar:2010we,Kostelecky:2016kfm}
Constraints on a number of parameters controlling modifications to the GW dispersion relation improved by factors of ${\sim}2.6$ with respect to GWTC-1 (Fig.~\ref{fig:liv}).\cite{Abbott:2020jks}
Moreover, this also resulted in a constraint of the graviton mass to $m_g \leq 1.76\times10^{-23}~\mathrm{eV}/c^2$, which is $1.8$ times more stringent than current Solar System bounds.\cite{Bernus:2020szc}
Details on these and other results, including consistency tests\cite{Ghosh:2017gfp} and measurements of the spin-induced quadrupole moment in BBHs,\cite{Krishnendu:2017shb} are outlined in Ref.~[\refcite{Abbott:2020jks}].

\section{Conclusion}
\label{sec:conclusion}

Although still a young field, GW astronomy has developed at neck-breaking speeds: little over five years after the groundbreaking detection of GW150914, LIGO and Virgo have now accumulated close to 50 compact-binary detections.
Individually and as a whole, these signals have provided invaluable insights about the nature and population of compact objects, galvanizing several subfields of physics and astrophysics.

During the first half of the third LIGO-Virgo observing run, there were on average ${\sim}1.5$ new compact-binary detections per week (Sec.~\ref{sec:det}).
These new discoveries included some unexpected sources: a pair of black holes with unequal masses (Sec.~\ref{sec:gw190412}), a massive pair of neutron stars (Sec.~\ref{sec:gw190425}), a black hole potentially in the supernova pair-instability mass gap (Sec.~\ref{sec:gw190521}), and either the lightest black hole or the heaviest neutron star known to date (Sec.~\ref{sec:gw190814}).
Collectively, the set of GWTC-2 signals revealed nontrivial features in the black-hole mass spectrum, provided evidence of spin precession, and confirmed previous signs that the BBH merger rate is likely to increase with redshift (Sec.~\ref{sec:astro})---all of which informs astrophysical population models, eventually improving our understanding of structure formation in the universe.
GWTC-2 also enabled more constraining and diverse tests of general relativity (Sec.~\ref{sec:tgr}), yielding factors of ${\sim}2$ improvements on the measurement of coefficients controlling GW generation and propagation, and allowing for new probes of the fundamental nature of black holes.

Data from the second half of the third LIGO-Virgo observing run are currently being analyzed---if history is any indication, those results are guaranteed to contain many new surprising discoveries.
At the same time, the instruments are scheduled for further updates, which will continue to increase their sensitivity in preparation for their fourth observing run.%
\footnote{The operation schedule has been impacted by the COVID-19 pandemic and is currently in flux; at the time of writing, the fourth observing run is scheduled to begin no sooner than June 2022.\cite{schedule}}
Furthermore, detections will soon be enhanced once the Japanese observatory KAGRA\cite{Somiya:2011np,Aso:2013eba} becomes fully operational, contributing to the sky-localization accuracy and duty cycle of the global GW detector network.\cite{Aasi:2013wya}
All these advances are certain to make of GW astronomy a fruitful and dynamic field for years to come.

\section*{Acknowledgments}
M.I.\ is supported by NASA through the NASA Hubble Fellowship
grant No.\ HST-HF2-51410.001-A awarded by the Space Telescope
Science Institute, which is operated by the Association of Universities
for Research in Astronomy, Inc., for NASA, under contract NAS5-26555.
LIGO was constructed by the California Institute of Technology and
Massachusetts Institute of Technology with funding from the National
Science Foundation and operates under cooperative agreement PHY-0757058.
This paper carries LIGO document number \dcc{}.

\bibliographystyle{ws-mpla}
\bibliography{refs}
\end{document}